# A Survey on Web-based AR Applications

BEHRANG PARHIZKAR[1], ASHRAF ABBAS M. AL-MODWAHI[2], ARASH HABIBI LASHKARI[3], MOHAMMAD MEHDI BARTARIPOU[4], HOSSEIN REZA BABAE[5]

**Faculty of ICT, Limkokwing University of Creative Technology,
Cyberjaya, Selangor, Malaysia**

[1]*hani.pk@limkokwing.edu.my*
[2]*al-modwahi@live.com*
[3]*a_habibi_l@hotmail.com*
[4]*mehdi.bartari@limkokwing.edu.my*
[5]*babaei@limkokwing.edu.my*

**Abstract**
Due to the increase of interest in Augmented Reality (AR), the potential uses of AR are increasing also. It can benefit the user in various fields such as education, business, medicine, and other. Augmented Reality supports the real environment with synthetic environment to give more details and meaning to the objects in the real word. AR refers to a situation in which the goal is to supplement a user's perception of the real-world through the addition of virtual objects. This paper is an attempt to make a survey of web-based Augmented Reality applications and make a comparison among them.

## 1. Introduction

Augmented Reality (AR) refers to a situation in which the goal is to supplement a user's perception of the real-world through the addition of virtual objects [1]. Due to the increase of interest in Augmented Reality, the potential uses of AR are increasing also. AR can benefit the user in various fields such as education, business, medicine, and other. The concept of enhancing persons perception of reality dates back to the 13th century when Roger Bacon made the first recorded comment on the use of eye glasses, for optical purposes. In 1665, an experimental scientist Robert Hooke introduced an idea of augmented senses in his book Micrographia. Ever since fiction writers, military industry and lately academic and commercial researchers have paved the road for augmented reality with an increasing effort.

1.1 AR in the virtuality continuum

Milgram on 1994 used the reality-viruality continuum to introduce a taxonomy that relates augmented reality to virtual reality as different degrees of reality-virtuality continuum [2].

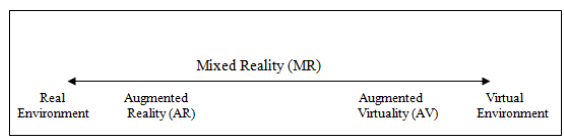

Figure 1: Simplified representation of a "virtuality continuum"

In the left end of the virtuality continuum is the real environment. An entirely immersive virtual environment is in the other end. Augmented reality is near the real environment end, as it consists of some synthetic elements that overlap the actual real environment. The inverse case where real world content contributes to synthetic surroundings would be called augmented virtuality.

1.2 Desktop AR

Seokhee on 2006 examined the comparative usability among three different viewing configurations of AR system that use a desktop monitor instead of a head mounted display (HMD). They found that, placing a camera in the back of the user was the best choice for convenience and attaching a camera on the user's head for task performance. The results was a guide for designing desktop augmented reality systems without head mounted displays [3]. They used a set of experiments and questioner to perform their investigation. From the analysis data, the best option was the head camera condition among the three tested in terms of task performance, there was no significant difference in task performance between the head mounted camera and the back camera conditions, all the results consistently rated the head mounted case the best in terms of task performance and most of the usability questions. But taking into account the effort of the system set up and the HMD cost, the second viewing condition (back camera) is also a good choice for the AR system without HMD. The back camera deliver the same level of task performance as the condition when the camera was mounted on the user's head and scored reasonably across the usability questions. The term "Desktop AR" is in fact more commonly used to indicate that the desktop is used as the interaction space. In short, the viewing condition depend on the purpose of the AR system itself.

### 1.3 Mobile AR

Anders Henrysson on 2005 described how mobile phones are an ideal platform for augmented reality and how they can also be used to support face to face collaborative AR gaming.They create a custom port of the ARToolkit library to the Symbian mobile phone operating system and then developed a sample collaborative AR game based on this. They also provide general design guidelines that could be useful for mobile AR applications developers [4]. After two experiments applied to the sample collaborative AR game and surveys . They found that, subjects would often grasp the cell phone with both hands and start intently at the screen while playing, never looking across the table at thier partner. Although they were collaborating in a face to face setting the focus of their attention was on the small screen. They also found that, each of the three conditions provides less visual information about the player's partner.These results show that users do feel that multi-sensory output is indeed important in face to face AR gaming. They almost unanimously rated the condition which provided the most sensory output (audio, visual, haptic) as easiest to work in and also as the most enjoyable.

### 1.4 Web AR

(Jens de Smit, 2010) indicated how AR applications use some parts of web infrastructure standards and how they use some proprietary technologies as well becuase the suite of web standards doesn't offer everything that is required to build AR applications.He discussed what would be required to be able to build web- based AR applictions  and how to prepare the suite of web standards for future developments in ubiquitous computing technology [5]. Jens provides examples to show how some devices such as iPhone and Android used  web technology in their native applications.AR browsers Layar and Wikitude use JSON and XML respectively to transport thier store and exchange points of interest (POIs).Flash and other web plug-in also used  for the web-based AR desktop applications.

 (Mohit Virmani et al., 2010) described how epipolar geometry, homography and fundametal matrix estimaiton can be used to rebuild the gap between AR and the web. Also they  present some of the potential problems with web and mobile AR [6]. They stated that, the current methods of optical or video technology which are being used for AR do not employ the provision of the real time access and usage of online available databases. The result was a system that employ the provision of the real time access to the online database.Also they found a solution for some issues in the web AR such as the security and spam ; Individual Augmented ID need to develop and the initiation of effective privacy management solutions including hardware, software, standards, and legal frameworks.

## 2. Web-based AR applications

There are various types of applications based on AR on the World Wide Web.Most of them are  web-base applications.

### 2.1. Web-based AR in education
#### 2.1.1. MARIE

Fotis Liarokapis on 2002 presented their application for engineering education, which is an interactive multimedia augmented reality interface for e-learning (MARIE). It developed in order to enhance traditional teaching and learning methods, MARIE is equally applicable to other areas. The authors have developed and implemented a user-friendly interface to experimentally explore the potential of AR by superimposing virtual multimedia content (VMC) information  in an AR tabletop environment, such as a student desk workspace.Users can interact with the VMC, which is composed of three dimensional objects, images, animations, text (ASCII or 3D) and sound [7].

They provide the feasibility of the system, only a small part of the teaching material was digitised and they presented some experimental results see (Figure 2) .

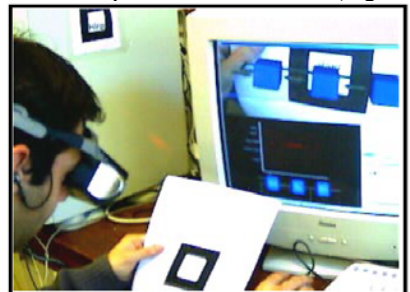

Figure 2: Monitor augmented reality view

On the monitor, the main window represents the augmented world part, where the 3D objects are superimposed on the tabletop environment in real time.also the 3D text is overlaid on the window; it's acting as help option for the user. The bottom left window plays an animation describing the object and the theory.The bottom right window displays images representing 2D diagrams.

### 2.1.2. Scimorph

Scimorph is a web-base AR education application for children. Scimorph is a central character who can journey through a series of activities in a virtual science environment based around the curriculum for primary aged children. Scimorph can be used at school or in the home to build on knowledge and understanding of the world. He has some human characteristics and attributes though he has not developed fully because of his lack of understanding about the world. Scimorph provides opportunities to discuss and solve scientific based problems, take part in discussions around the activities and delve deeper into the topic by means of interactive tools and use of web based materials [8]. The figure below shoes the website of the application. It contains three main buttons the first one to print the marker, the second to download the guidance notes and the last to try it out.

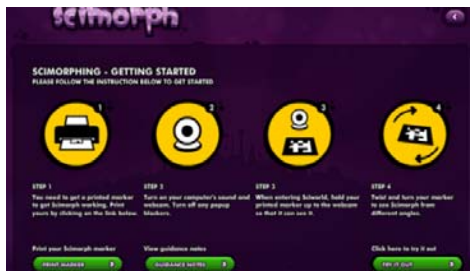

Figure 3 : screen shot for scimorph.

### 2.1.3. LearnAR

LearnAR is an AR web-based application for e-learning , it is a powerful learning tool that brings investigative, interactive and independent learning to life using AR . It is a pack of ten curriculum resources for teachers and students to explore by combining the real world with virtual content using a webcam. The resource pack consists of interactive learning activities across English, maths, science, RE, physical education and languages that bring a wonderful factor to the curriculum [9].

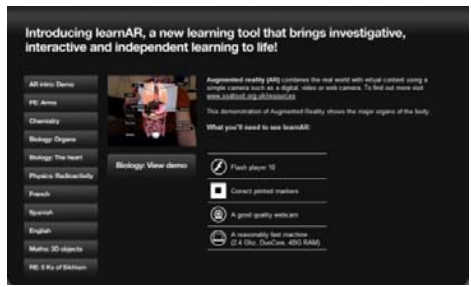

Figure 4 : screen shot for learnAR

The figure above shows the interface of the application on its website, it provides some instructions and conditons to be able to use the application. The application works fast because it is based on flash.

## 2.2. Web-based AR in medicine
### 2.2.1. Surgical AR system

J. Fischer on 2004 proposed in a paper an alternative approach of building a surgical AR system by harnessing existing, commercially available equipment for image guided surgery (IGS). They provide a detailed report of the prototype of an augmented reality application, which receives all the important information from a device for intraoperative navigation [10].

They have implemented an application demonstrating the feasibility of IGS based medical AR, several images generated by the software see figure 5. After they test their application the result was a system capable of generating an augmented video stream at average frame rates of more than 10 fps using a webcam resolution of 640 x 480 pixels.They have measured an average latency of approximately 80 ms for receiving the tracked instrument data from the Vector Vision system. Their camera tracking method was good in terms of accuracy, but sporadic visual mismatches result from the time lag between IGS marker registration and the generation of the augmented video. The range of the tracking methods is limited to the viewing volume of the infrared cameras, and it is sensitive to occlusion of the marker clamps from their viewpoint. This make the application more ill-suited for the use of HMD.

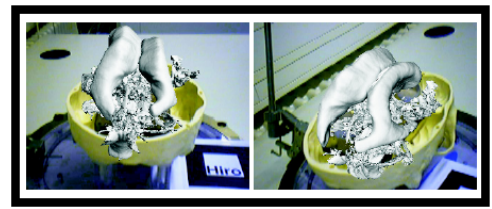

Figure 5 : Several views generated by the AR system using a ventricle model generated from a MRI scan.

### 2.2.2. Anatomy AR learning system

Chien-Huan Chien on 2010 examined in a study the possibility of using AR to create an interactive learning system, which helps medical students to understand and memorize the 3D anatomy structure easily with tangible augmented reality support. They speculate that by working directly with 3D skull model with visual support and tangible manipulate. The complex anatomy structure can be learned faster and better with thier system [11]. The system is based on a complete structure of the skull which can be decomposed and reassembled. To be an effective training tool, it has to provide correct information to the students, the skull includes zygomatic bone, temporal bone, sphenoid bone, mandible, maxilla, ethmoid bone, parietal bone, frontal bone, occipital bone, nasal bone, lacrimal bone, palatine, vomer, and inferior nasal concha. With a clear pop-up labeling and interactive 3D model, students can easily get the related position of each bone in different angle.In order to use the system, only a computer with webcam and a marker are needed. The purpose of the anatomy AR learning system is to examine how medical students learn and interact with a computer-garnered 3D skull in the augmented reality system that helps them to identify the related position of skull and to memorize the skull's structure.

### 2.2.3. AR system for medical training

Felix G. Hamza-Lup on 2009 presented a distributed medical training prototype designed to train medical practitioners' hand-eye coordination when performing endotracheal intubations. It accomplishes this task with the help of AR paradigms. By employing deformable medical models an extension of this prototype is possible [12].

The system will allow paramedics, pre-hospital personnel, and students to practice their skills without touching a real patient and will provide them with the visual feedback they could not otherwise obtain. It has the potential to allow an instructor to simultaneously train local and remotely located students and allow students to actually see the internal anatomy and therefore better understand their actions on a human patient simulator (HPS) see figure 6 . There are some issues arise in the design and implementation of these applications on a distributed systems infrastructure. The complexity of such systems triggers assessment difficulties, such as a complex issue is related to the distributed measurements taken for the ASA assessment.

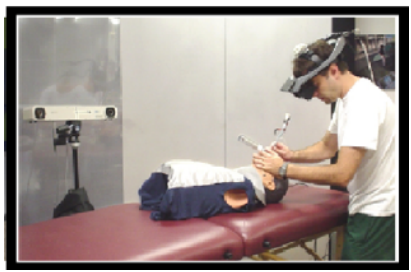

Figure 6 : Illustration of the AR tool for training paramedics on ETI

### 2.3. Web-based AR in game

(Christiane Ulbricht and Dieter Schmalstieg, 2003) attempt to demonstrate that tangible augmented reality is a highly effective environment for specific types of multiplayer computer games.They presented their prototype system through both theoretical argumentation and presentation to show how the usefulness of this user interaction paradigm.Also they evaluated several variants of the interaction techniques necessary for the game during the development [13].

Tangible Augmented Reality (TAR) is a combination of an Augmented Reality System and a Tangible User Interface [14]. A user interacts with virtual objects by manipulating real objects. A tangible user interface (TUI) uses objects of the natural enviroment as an interface instead of the computer interface [15].

TUI can be used by several persons because it is not restricted to one screen or one keyboard, see figure 7.

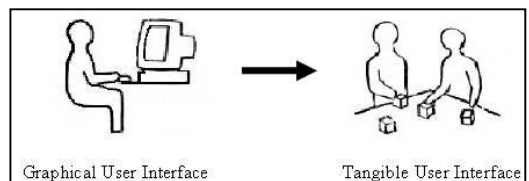

Figur 7: A Tangible User Interface uses objects of the natural environment as an interface to the computer.

Christiane and Dieter found some advantages of using TUI during experiment their tabletop game, the following are the advantages:
- Virtual objects become tangible.
- It is a wireless interaction.
- The amount of available input devices increases.

### 2.3.1. AR photography game

Cody Watts and Ehud Sharlin on 2008 presented a photography-based AR game, it's a game that involves two players, and each player uses a physical handheld camera device to take pictures of floating virtual ghosts. Players must creep, sneak, and maneuver themselves through physical space in order to approach their ghostly subjects and snap a picure using their paranormal camera [16].

Photogeist's gameplay is based on a combination of two everyday activities; physical locomotion and photography.In this game the developers have encouraged players to move as they play by creating ghosts wich are constantly moving and reacting to the player's presence, the ghosts will naturally seek to avoid the player.If the player stands still, she will find herself with very few opportunities to take compelling photos. Thus, if a player wishes to score highly, she will find it necessary to continually move and reposition herself throughout the game.Photography is a major part of this game, with the proliferation of digital cameras and mobile's camera in modern society, taking pictures has become an everyday skill. They are leveraging this common knowledge to create a game that is very intuitive for first-time players , see figure 8.

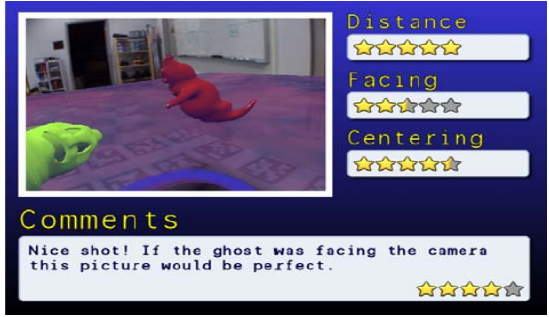

Figure 8: In-depth information on a photograph.

### 2.3.2. AR racing game

Fotis Liarokapis on 2009 presented a pervasive AR serious game that can be used to make better entertainment using a multimodal tracking interface. The main purpose of their research is to design and implement generic pervasive interface that are user-friendly and can be used by a side range of users including people with disabilities. A pervasive AR racing game has been designed and implemented. The goal of the game is to start the car and move around the track without colliding with either the wall or the objects that exist in the gaming arena. Users can interact using a pinch glove, a Wiimote, through tangible ways as well as through I/O controls of the UMPC. Initial evaluation results showed that multimodel-based interaction games can be beneficial in serious games [17].

### 2.4. Web-based AR in marketing
### 2.4.1. Ray-ban

Rayban has launched an advertisement using augmented reality in its website, which gives users the ability to try the glasses instead of going to the shops and spend much time for chosing and trying them [18]. The advertisements is an AR application must be connected to the internet to work, when the user download the application from Rayban website and open it, the application stored itself in the computer and then strart conect to a server.In order to use this application, minimum system requirements Windows® XP/ Windows Vista® OS - Intel Pentium® IV 1.5 Ghz or AMD Athlon XP® 1500+ Mhz - 1GB of RAM (1,5GB on Windows Vista®) - 64 MB 3D graphics card compatible DirectX® 9.0c (support shaders 3.0). DirectX® 9.0c (included) - 2.0 GB available HD space - A 56k or better Internet connection - A 640x480 compatible webcam. Note: The user must have administrator rights to the computer he is using. Recommended Specifications Windows® XP/ Windows Vista® OS - Intel Core 2 Duo or AMD Athlon X2 - 2GB of RAM - 256 MB 3D graphics card compatible DirectX® 9.0c (support shaders 3.0) - DirectX® 9.0c (included) - 2.0 GB available HD space. Broadband Internet connection - A 640x480 compatible webcam, the figure below shows how it looks.

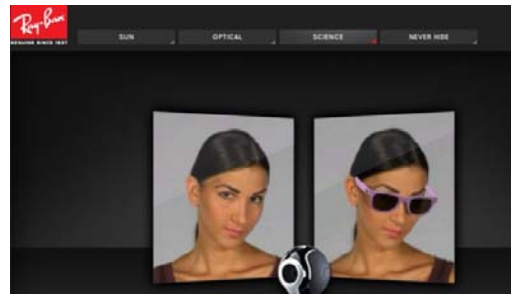

Figure 9 : screen shot for AR Rayban

### 2.4.2. Adidas

Chris Barbour, head of digital marketing for Adidas Originals says, "That's what we have done. We have taken a real world item and added a fantastic virtual world on top of that " [19]. Adidas launched an AR advertisement in its website, users have to hold up their new shoes to Adidas' website in order to access this magical fantasy world via a code embedded in the trainer's tongue.Users hen use their flashy sneaker as a controller to navigate their way around this world. To get started download a special code from adidas website for a sneak peek into the adidas neighborhood. If the users are wanting more, can head into any champs sports store and pick up a special code that will give them access to and exclusive star wars game, see figure 10 [20].

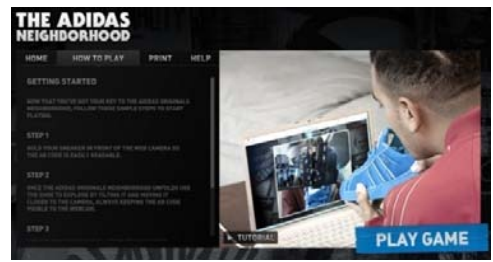

Figure 10 : screen shot for AR Adidas

### 2.4.3. Nike

Nike launched an AR advertisement which is an AR application that can be downloaded from nike website to enjoy watching its products.
Here are five easy steps users must follow:

1) Make sure you have a webcam hooked up to your computer.
2) If you've been fortunate enough to get your hands on the banging Dusty Payne or Mike Spinner advertisements that were featured in the US Open of Surfing or Dew Tour spectator guides, keep that handy because you'll need it.
3) Download the 6.0 augmented reality file to your computer.
4) Launch the Nike_6_0_AR.exe file that you just downloaded, and follow the on screen instructions to install the application. It should only take 1- 2 minutes to set up.
5) Once you initiate the application, all you need to do now is put the Dusty Payne or Mike Spinner ad in front of your webcam and watch the action come to life in 3-D (press 1,2,3 or 4 on your keyboard to change the videos). Figure 11 shows nike's marker which gives you the chance to win a free pair of shoes [21].

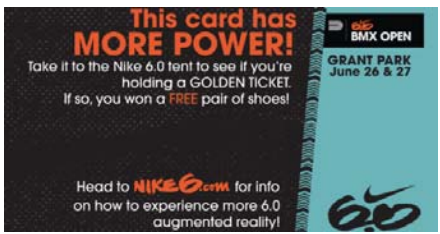

Figure 11 : screen shot for AR Nike

### 2.4.4. Tissot

Tissot creates an AR advertisement in its website, which makes users experience the Tissot Touch collection on their own wrist instead of going to shops and spend a lot of time for chosing the desired watch.
The following is what users need to start :
1) Cut around the wristband and along the horizontal red lines to enable you to wear around your wrist.
2) Wear the wristband around your wrist.
3) Turn on your webcam and download the software from www.tissot.ch/reality.
4) Hold your wrist in front of the webcam. You will see the T-Touch Expert watch appear. You can now try on the Touch collection and experience the different functions of the Tissot touch screen watches, see figure 12 [22].

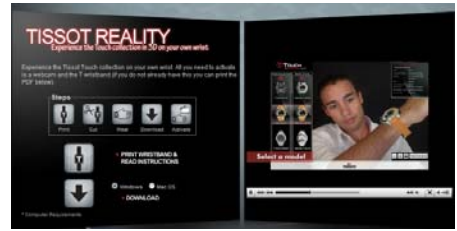

Figure 12 : screen shot for AR Tissot

### 2.4.5. Mini-cooper

MINI uses augmented reality technology to create a truly interactive media piece out of a 2-dimensional magazine ad. Using augmented reality tracking technology, as users hold the ad up to their computer's web cam, they'll see a 3-D model of a MINI Cooper convertible that moves as they turn and move the sheet of paper around It looks as if they are actually holding the MINI Cooper car in thier hands.

All you need is a copy of the printed ad (working as a grid), a web cam, and a web browser to view the 3-D Augmented Reality effect. A great way to create buzz around a product and to get viewers to go online and interact with the product.

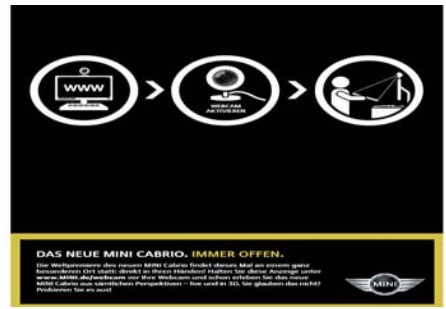

Figure 13 : screen shot for AR Mini

This is what the actual MINI Cooper print ad looks like and it appeared in three German automotive magazines; Auto Motor und Sport, Werben & Verkaufe and Autobild [23].

### 2.4.6. BMW

BMW initiated an online augmented reality campaign to promote the launch of the BMW Z4. Inition worked with dare to create a unique interactive online brand experience supporting the TV campaign in which a roadster is steered across a blank canvas with the tires providing paint trails and colourful, see the figure below.

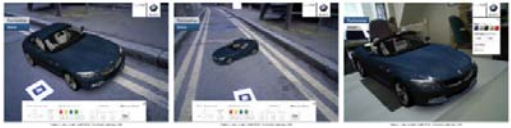

Figure 14 : screen shot for AR BMW

Powered by Inition's augmented reality technology MagicSymbol, users can now create their own "expression of joy" online. Users can test drive a virtual 3D Z4 and record their paint trails via the webcam and upload videos of their creations to YouTube. The augmented reality application is complemented by content on Facebook and YouTube.BMW has always taken a traditional approach to marketing and is trying to move more towards digital. Yet it's not a question of throwing out traditional media because obviously that's still important for the brand [24].

### 2.4.7. Toyota

Toyota have used augmented reality to create 3D a interactive experience of the new iQ car which users can download from toyota website. Toyota iQ is a radical new small car and augmented reality technology allows you to interact with the car to discover its agility and interior space.

To experience iQ Reality users need at least 1GB of spare RAM, a good graphics card and a webcam. PCs must be running Windows 2000, XP,Vista or Windows 7; Macs should be on OSX 10.4. Computers set up for gaming are ideal. iQ Reality begins when you show one of the two printed symbols to your webcam. The camera needs to read the symbol to generate the 3D image. It won't work if you are covering any part of the printed symbol. Glaring lights from overhead can sometimes interfere with the technology. If your webcam isn't reading the printed symbol properly, try holding a magazine or a piece of card underneath your print out, see figure 15 [25].

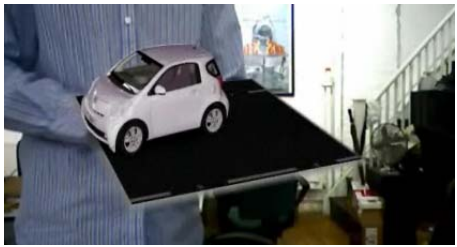

Figure 15 : screen shot for AR Toyota

### 2.4.8. Nissan

Nissan Australia have launched a very cool 370Z augmented reality website created by Tequila. Potential buyers have been sent a cool DM piece, complete with a steering wheel cut-out that becomes the receptor for the new Nissan 370Z augmented reality application (or users can just print one). All you need is a web cam, the steering wheel and visit Nissan website, see figure 16 [26].

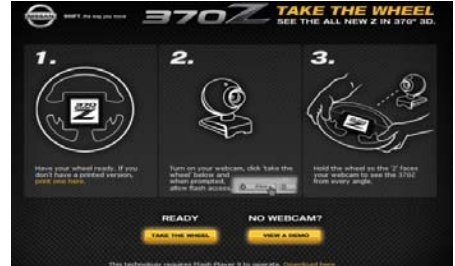

Figure 16: screen shot for AR Nissan

Its a pretty cool little experience, where you have control over the car directly in your hands, can set it on a showroom type spin or just move it around to check out all the details, it even has a virtual "book a test drive" button. Still not as amazing as the BMW Z4 augmented reality website, but probably the first Australian agency to do this for an automotive brand [27].

### 2.4.9. Volvo

Adv.pl is an advertisement company that is just implementing the first in Poland augmented reality project in the automotive industry . This project supports launching of new Volvo S60. They were responsible for creating augmented reality technology for interactive booklet showing greatest merits on the new Volvo S60 model. Each of four booklet pages uses other capabilities of augmented reality: 3D presentation of the new car, virtual simulation of vehicle colours available in the market and a film presenting operation of Pedestrian Detection security system that consists in recognizing moving objects on car route. The agency created a realistic 3D video model of a new car on the basis of photos and available visualizations [28].

A booklet based on augmented reality technology will be distributed in 19 showrooms of Authorized Volvo Dealers in the whole Poland, where the customers will be able to enter the world of augmented reality and watch closely the car. Adv.pl are the only company in Poland that obtained exclusive licence rights from French Total Immersion company (leader in the global Augmented Reality market) for delivering advanced Augmented Reality to the Polish customers.The figure below shows how the advertisement looks like.

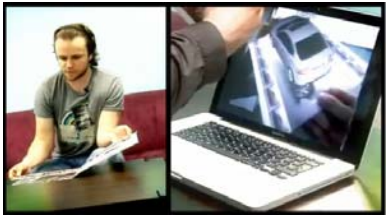

Figure 17 : screen shot for AR Volvo

## 3. Discussion

After reviewing the above AR applications, an comparison will be applyed among the applications in terms of the following:

### 3.1. Quality

The quality of an application measures how well appliction is designed, and how well the application conforms to that design.

### 3.2. Simplicity

Simplicity is a more qualitative word connected to simple. It is a property, condition, or quality which things can be judged to have. It usually relates to the burden which a thing puts on someone trying to explain or understand it. Something which is easy to understand or explain is simple, in contrast to something complicated. In some uses, simplicity can be used to imply beauty, purity or clarity.

### 3.3. Usability

Human-Computer-Interaction (HCI) is the area where usability emerged. Several books or papers about HCI present a definition or characterization of usability. For instance,(Hix & Hartson,1993) consider that usability is related to the interface efficacy and efficiency and to user reaction to the interface.

### 3.4. Efficiency

The ratio of the output to the input of any system.It also refers to skillfulness in avoiding wasted time and effort.

### 3.5. Availability

The degree to which a system, subsystem, or equipment is operable and in a committable state at the start of a mission, when the mission is called for at an unknown, availability is the proportion of time a system is in a functioning condition.

Table 1, shows an evaluation table among all presented applications based on these attributes.

## 4. Conclusion

This paper explained about AR, its importance and how it can benfit the user in various fields such as education, medicine, and marketing etc., This paper observed AR web-based applications, and compared among them. According to the table above the best applictions are Ray-ban, Tissot, and BMW.

Table 1: Comparison table among AR applications

| | Quality | Simplicty | Usability | Efficiency | Availability |
|---|---|---|---|---|---|
| MARIE | | √ | | | |
| Scimorph | √ | | | √ | √ |
| LearnAR | √ | | √ | √ | |
| Surgical AR system | | √ | | | |
| Anatomy AR learning system | | √ | √ | | |
| AR system for medical training | √ | | √ | | |
| AR photography game | √ | | √ | √ | |
| AR racing game | | √ | | | |
| Ray-ban | √ | √ | √ | √ | √ |
| Adidas | | √ | | | √ |
| Nike | | √ | | | √ |
| Tissot | √ | √ | √ | √ | √ |
| Mini-cooper | √ | √ | | | |
| BMW | √ | √ | √ | √ | √ |
| Toyota | √ | √ | √ | | √ |
| Nissan | | √ | √ | | √ |
| Volvo | | √ | | | √ |